# Design and Fabrication of Acoustic Wave Actuated Microgenerator for Portable Electronic Devices

[1]Tenghsien Lai, [2]Changhan Huang, and [2]Chingfu Tsou

[1]The Graduate Institute of Electrical and Communications Engineering, Ph.D. Program,
Feng Chia University, 100 Wenhwa Road, Seatwen, Taichung, Taiwan, 40724 R.O.C.
[2]Department of Automatic Control Engineering, Feng Chia University,
100 Wenhwa Road, Seatwen, Taichung, Taiwan, 40724 R.O.C.,

*Abstract-* An acoustic wave actuated microgenerator for power system applications in mobile phone was design, fabricated, and characterized. We used the acoustic wave of human voices or speakerphone by way of an electromagnetic transducer to produce electrical power for charging or to run a portable electronic device. The proposed microgenerator is composed of a planar coil, a suspension plate with supporting beams and a permanent magnet. In this paper, the dynamic response of the suspension structure and the variation of the induced magnetic flux were characterized by using commercial finite element analysis and Ansoft Maxwell EM3D software, respectively. The electroplating nickel and silicon bulk micromachining techniques were used to fabricate the suspension plate and planar coil, and by integrating a permanent magnet as well as wafer to wafer adhesion bonding to accomplish the microgenerator assembly. The experimental results shown the typical microgenerator with a planar dimension of 3 mm × 3 mm, the maximum induce-voltage 0.24 mV was generated at the driving frequency of 470 Hz. The dynamic response of microgenerator can be designed to meet a specific acoustic driving frequency to increase the efficiency of energy harvesting.

## I. INTRODUCTION

The past few years have seen an increasing focus on energy harvesting issue, including power supply for portable electric devices. Utilize scavenging ambient energy from the environment could eliminate the need for batteries and increase portable device lifetimes indefinitely. Several different ambient sources, including solar, vibration and temperature effect, have already exploited [1-3]. In particular, scientists are currently trying to find more efficient ways of utilizing solar cells for electricity production [4-6]. Each energy source should be used in suitable environment, therefore to produce maximum efficiency.

Ambient energy generators have the potential to replace battery power as an energy source in a variety of practical applications. For example, kinetic (or vibration) energy generators harvest electrical energy from movement present in the application environment. Various kinetic generators use piezoelectric materials [7~10], electrostatic [11~13] and electromagnetic [14~20] transduction mechanisms. Piezoelectric materials are perfect candidates for vibrational energy scavenging as they can efficiently convert mechanical strain to an electrical charge without any additional power, but it is difficult to implement at micro-scale. Electrostatic is a good solutions for the construction of microgenerators because of its relative ease of integration with microelectronics. The disadvantage of this approach is that a separate voltage source is required to charge the capacitor. Electromagnetic conversion uses vibration to move a conductor in a magnetic field. Existing prototypes generate low voltage output to be usable. However, the amount of energy generated by those approaches depend the application environment and the efficiencies of the generators and the power conversion electronics. The goal is to develop miniature, self-contained and renewable power supplies that convert energy from an existing source in the environment into electrical energy.

In this study, we present an acoustic wave actuated microgenerator with high output voltage suitable for commercial portable electronic devices. We use the energy of acoustic waves, such as the sound from human voices or speakerphone, to actuate a MEMS-type electromagnetic transducer. This provides a longer device lifetime and greater power system convenience. It is convenient to integrate MEMS-based microgenerators with small or portable devices. Our aim is to run a small electronic device, like a cell phone, and may soon be able to power a laptop.

## II. DESIGN AND ANALYSIS

The proposed acoustic wave actuated microgenerator, as shown the Fig. 1, consists of a top wafer with planar coil and a bottom wafer with a suspended flat and support beams as well as a high magnetic energy permanent magnet. The top wafer is etched through its thickness by anisotropic wet etching (KOH). The wafer-to-wafer bonding process is aligned to ensure correct placement of the coil relative to the magnet. The top surface of the magnet is in line with the planar coil. The thickness of the magnet is same as that of the top wafer. Thus, the planar coil has tight gap spacing with the permanent magnet; this enhances the induced voltage. Larger output power can be obtained by link an array of identical microgenerators.

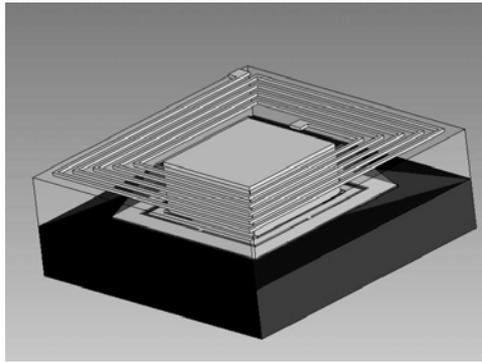

(a)

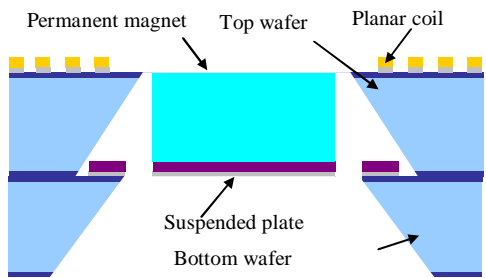

(b)

Figure 1. Design concept of the microgenerator: (a) 3-D and (b) Cross-section view.

Table 1 Material and geometry parameters of microgenerator for analysis process

| Material parameters | Value | Scale |
|---|---|---|
| Young's modulus (E) | $2\times10^{11}$ | Pa |
| Structure density ($\sigma_1$) | 8910 | kg/m$^3$ |
| Magnet density ($\sigma_2$) | 9000 | kg/m$^3$ |
| Geometry parameters | Value | Scale |
| Beam length (L) | 800 | µm |
| Beam width (W) | 60 | µm |
| Beam thickness (H) | 20 | µm |
| Flat length ($X_1$) | 2000 | µm |
| Flat width ($Y_1$) | 2000 | µm |
| Flat thickness ($Z_1$) | 20 | µm |
| Magnet length ($X_2$) | 2000 | µm |
| Magnet width ($Y_2$) | 2000 | µm |
| Magnet thickness ($Z_2$) | 500 | µm |

The commercial software ANSYS was used to simulate the resonant frequency and vibration mode of suspension microstructure for the actuation frequencies produced by human vocal cords. The purpose is to achieve the micro-generator with higher charge collection efficiency in a small area under a specific driving frequency. Noting the frequency range of 200 Hz - 1.5 kHz in sound pressure level characteristics is the most important range in human vocal and electrical acoustic instruments, as shown in Fig. 2. The schematic diagram of the mechanical structure is shown in Fig. 3(a). The middle point of each suspension beam is fixed to the silicon substrate. The material parameters and device dimensions of the microgenerator are shown in table 1. The typical simulated result for the suspended flat with a permanent magnet is shown in Fig. 3(b). The figure demonstrates the first resonance mode of the microstructure is an out-of-plane piston-like motion and the natural frequency is 1012 Hz.

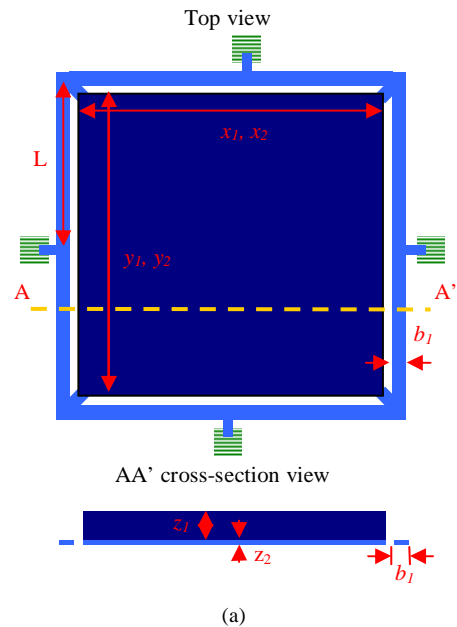

(a)

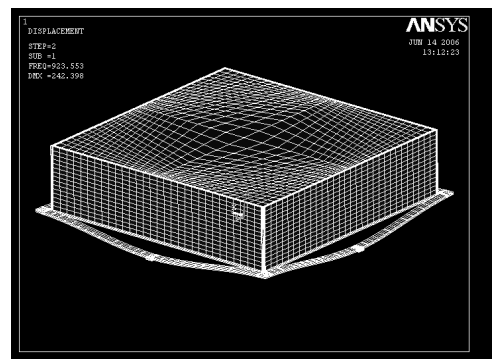

(b)

Figure 3. (a) The design of the mechanical structure and (b) its first modal (piston-like motion) of vibration that simulated by ANSYS

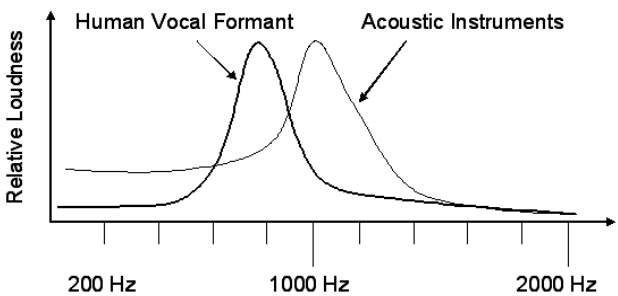

Figure 2. Frequency responses of human vocal formant and commercial acoustic instruments

The electromagnetic software Ansoft Maxwell EM3D was employed to characterize the variation magnetic flux due to the mechanical and electrical energy conversion is coupled through the magnetic field across the conductor coil. In this simulation process, we design fifteen-turn square spiral coil made from Cu with 20 µm gap spacing on a substrate. Each coil has a width of 20 µm and a thickness of 10 µm. The first-turn coil has one winding over the edges of the magnet. The size of magnet is same as in table 1. Figure 4 show the typical simulation result for the magnet above one-turn planar coil at a distance of 10 µm. The figure shows that the bottom edge of magnet closer to the coil, the more magnetic flux is induced. The magnetic flux drops very rapidly as the magnet moves away from the center of the coil. By the post process function of the Maxwell EM3D, the magnetic flux of each planar coil was calculated. For a 3 mm × 3 mm microgenerator, the estimated output voltage is approximately 0.58 mV at 1 KHz driving frequency with ± 50 µm amplitude range.

### III. FABRICATION

To meet the requirements of high-voltage output under a specific range of driving frequency, we used electroplating and bulk micromachining to fabricate the planar coil and suspension microstructure, and then integrating an Nd-Fe-B based permanent magnet on the suspended plate. The fabrication processes of the planar coil and suspension microstructure are shown in Fig. 5 and Fig. 6, respectively. Figure 5 shows the fabrication processes of the planar coil that starts with a 500 µm thick 4-inch <100> silicon wafer (Fig. 5a). First, the (100) single crystal silicon substrate was placed in a furnace at 1050 °C for 150 minutes to grow a 1 µm thick wet thermal oxide layer (Fig. 5b). Then a 0.1 µm-thick Cr film and a 0.2 µm-thick Ni film were deposited by thermal evaporation as adhesion and seed layers respectively (Fig. 5c). Next, a 25 µm-thick of photoresist AZ4620 was spun and patterned by photolithography. After that, a Ni layer was deposited by electroplating to fabricate the planar coil, and then the residue photoresist was removed (Fig. 5d). It is noted that the Cu with the excellent conductivity is suitable as the coil material than the Ni, however the process of electroplating Ni is easy to verify the possibility of our design and to improve the yield. To isolate the electrical properties, the exposed region of Cr adhesion and Ni seed layers were etched by a wet etch solution (Fig. 5e). The backside oxide layer was patterned by photolithography and then etched by BOE solution (Fig. 5f). After that, the exposed regions of silicon were wet etched by KOH solution (25%) at 80 °C for 4 hour to through the wafer thickness, and finally SiO2 layers were etched by HF to obtain a single layer of Ni microstructure (Fig. 5g).

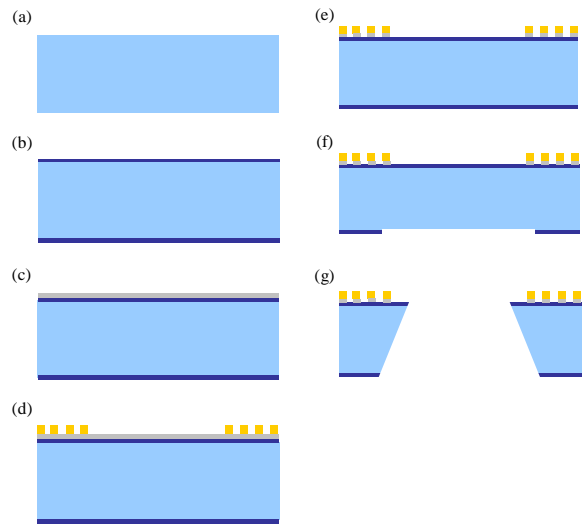

Figure 5. Fabrication process of planar coil

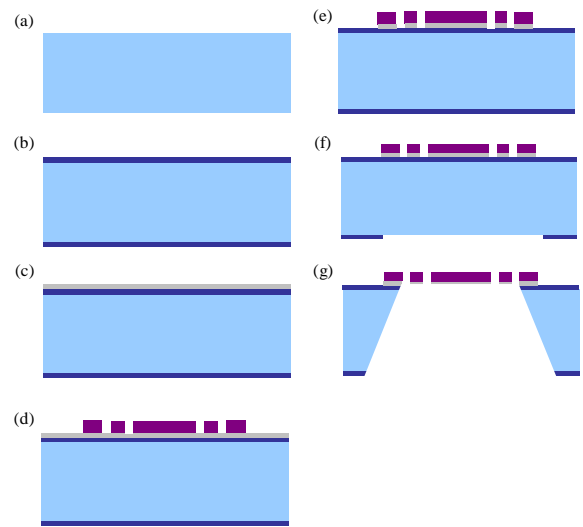

Figure 6. Fabrication process of suspension microstructure

The fabrication process of the suspension microstructure in Fig. 6(a) ~ (g) is same as that of the planar coil except step (d) and the mask layout. The Ni suspension microstructure is deposited by electroplating technology and is 20 µm thick. The electroplated Ni film has the advantages of low-residual stress, high density, and superior mechanical strength and can be formed as a high-aspect-ratio microstructure by photolithography [21-23]. As to the mechanical properties of electroplated Ni film from the previously works [24-27] referred and the quasi-static bending beam test using nanoindentation system in our research [28] presented are summarized in Table 2.

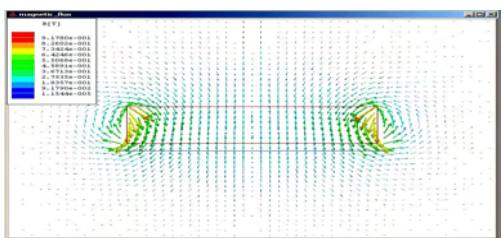

Figure 4. The magnetic flux in the space simulated by Ansoft MaxwellEM3D

Table 2 Mechanical properties of electroplated Ni film.

| References | Young's modulus E (GPa) | Yield strength σy (MPa) |
|---|---|---|
| [11-14] | 125~231 | 323~1550 |
| Our research presented [15] | 179~225 | 660~1120 |

## IV. Experimental Results

This research has successfully designed and fabricated an acoustic wave actuated microgenerator using the presented processes. Typical fabrication results for the planar coil and suspended plate are shown in Fig. 7 and Fig. 8, respectively. Figure 7 show the square spiral coil has fifteen-turn with 20 μm gap and each planar coil has a width of 20 μm and a thickness of 10 μm. The measured resistance from pads is approximate 58 Ω. The size of etching through-hole is 2 mm × 2 mm. Figure 8 shows the suspension microstructure with slight bending deformation from the fix end is due to the effect of residual stresses. This result would not affect the latter assembly and performance measurement of microgenerator significantly. Fig. 9 shows the top view of a 2 mm × 2 mm × 0.5 mm microgenerator with a permanent magnet. The pad setting is for the wire bond to output the induced electrical energy. The residual magnetic flux density (Br) of the magnet is 1.2 Tesla, according to the manufacturer's data.

To understand the dynamic characteristic and the performance of the microgenerator, Laser Doppler Vibrometer (LDV) and an acoustic speaker with a power of 3W were employed to measure the resonance frequency and induced-voltage of the microgenerator, as shown in Fig. 10. The functions of the power supply and oscilloscope are to provide the initial bias-voltage to detect the output voltage, respectively. The gap spacing between speaker and microgenerator was 5 mm. After a driving voltage of 1V (peak to peak) with sin wave signal has been input the speaker, the dynamic responses measured at atmospheric pressure by LDV is show in Fig. 11(a) and 11(b). Fig. 11(a) shown the magnet had a maximum displacement 2.8 μm at the first resonance frequency (piston-like motion) of 470 Hz.

Also, if the driving voltage is increased to 10 V, a maximum out-of-plane displacement of 11.5 μm can be achieved, as shown in Fig. 11(b). In this case, the measured induce-voltage at the first resonance frequency 470 Hz is 0.24 mV (peak to peak) when the driving voltage is 1 V. Therefore, using the proposed fabrication methodology and operation principle, we are able to manufacture a microgenerator with high-power output. The generated energy can apply to an electric cell or as a secondly battery, although the measured amount of electrical energy could not supply enough power to run a cell phone. In addition, high induced-voltage can be achieved by using the coil with high conductivity such as Cu or under a specific driving frequency by increasing the thickness and the number of coils, as well as by reducing the gap spacing between coil and permanent magnet.

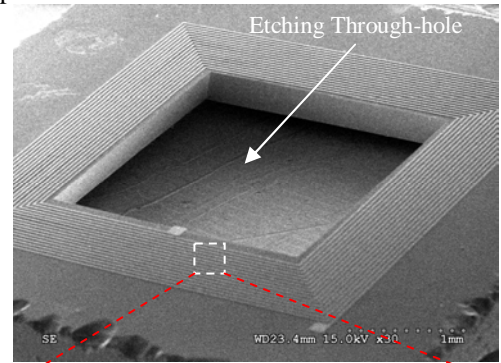
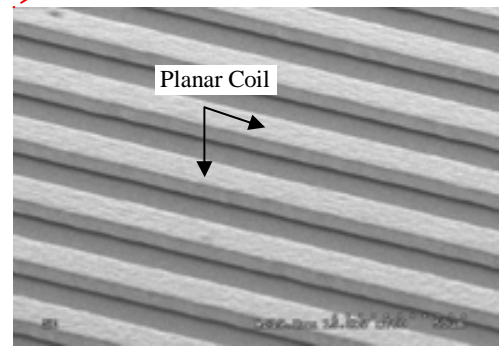

Figure 7. SEM image of the top wafer with planar coil

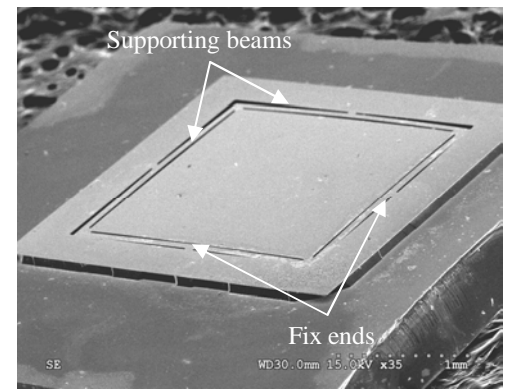

Figure 8. SEM image of the bottom wafer with suspended microstructure

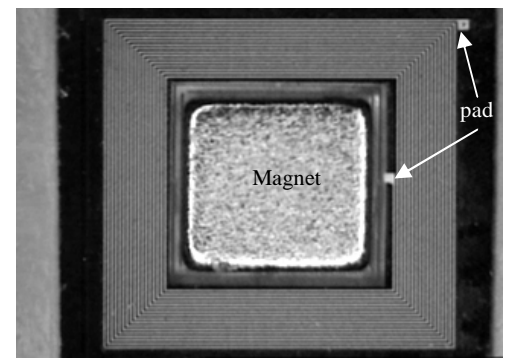

Figure 9. Top view of optical microscopy image of the assembled microgenerator

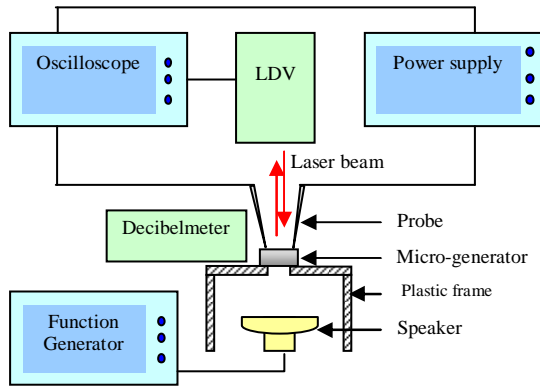

Figure 10. Experiment set up for the measurement of dynamic characteristic of microgenerator

electroplated Ni is about 14 µm, which is less than the 20 µm in the proposed design. In addition, the measured amplitude is only ± 3 µm, also less than the intended ± 50 µm significantly. However, we can predict the performance of the microgenerator accurately by using available data from past experiments and eliminating all errors associated with critical device dimensions to meet some specific acoustic driving frequency.

## V. CONCLUSION

A novel configuration for an electromagnetic microgenerator has been modeled and confirmed by experiments at a small macroscopic scale. The performance of the planar coil and suspension structure has been characterized by software analysis and experimental results. Fabrication result show a 3 mm × 3 mm× 1 mm microgenerator had maximum induce-voltage 0.24 mV obtained at a driving frequency of 470 Hz. Based on a MEMS type electromagnetic transducer, our design provides several important features such as the planar coil has the same level and a small gap spacing with the permanent magnet in in-plane to enhance the induced voltage. The suspension microstructure can be modified appropriately to meet a specific acoustic driving frequency to increase the efficiency of energy conservation. The small size, high efficiency, and potentially low cost make microgenerator desirable for numerous portable electronic devices. In the future, microgenerators can be combined with batteries and integrated circuit to make energy harvesting system for small devices.

ACKNOWLEDGMENT

This project was supported by the National Science Council of Taiwan under grant of NSC-95-2221-E-035-016. The authors appreciate the Precision Instrument Support Center of Feng Chia University, and the NSC National Nano Device Laboratory (NDL) in providing the fabrication facilities.

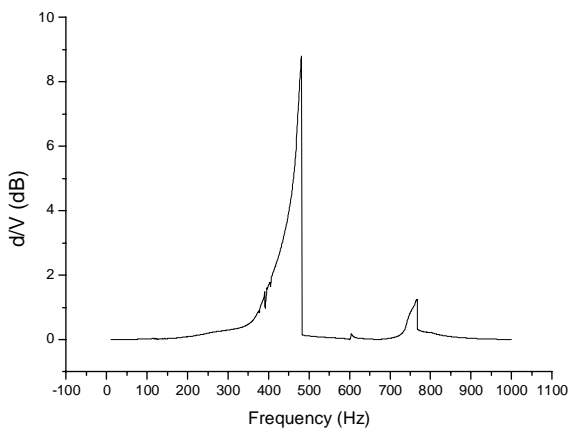

(a)

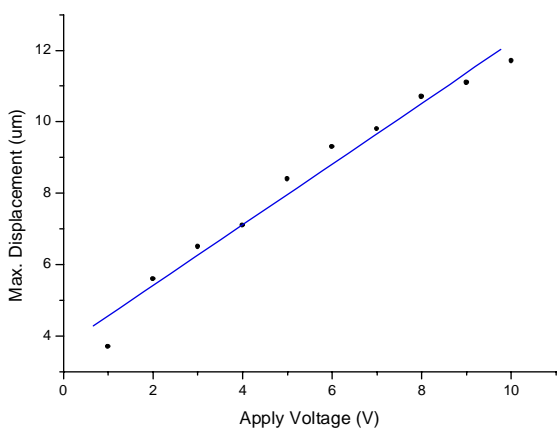

(b)

Figure 11. Measurement results of microgenerator: (a) the dynamic response illustration at the apply voltage of 1 V and (b) the out-of-plane displacement vs. applied voltage at the resonant frequency of 470 Hz.

Comparing experimental results to analytic predictions, we note many differences in first natural frequency and induced-voltage. We suggest many differences arise from the error in material parameters and in manufacturing geometry precision, the effect of the thickness of suspension structure may have been particular disruptive. The measured thickness of